**Away from Trolley Problems and Toward Risk Management**

Noah J. Goodall, Ph.D., P.E.
Virginia Transportation Research Council
530 Edgemont Road
Charlottesville, VA, USA 22903
noah.goodall@vdot.virginia.gov

**Abstract**

As automated vehicles receive more attention from the media, there has been an equivalent increase in the coverage of the ethical choices a vehicle may be forced to make in certain crash situations with no clear safe outcome. Much of this coverage has focused on a philosophical thought experiment known as the "trolley problem," and substituting an automated vehicle for the trolley and the car's software for the bystander. While this is a stark and straightforward example of ethical decision making for an automated vehicle, it risks marginalizing the entire field if it is to become the only ethical problem in the public's mind. In this chapter, I discuss the shortcomings of the trolley problem, and introduce more nuanced examples that involve crash risk and uncertainty. Risk management is introduced as an alternative approach, and its ethical dimensions are discussed.

**The Trolley Problem**

A self-driving car is driving toward a tunnel when suddenly a child runs into the road from behind a rock. The car begins to brake, but its software determines that braking alone will not slow the car enough to stop in time, or even reach a survivable impact speed. The vehicle can, however, swerve while braking. The car has a decision: stay in the roadway and hit the child, likely killing him, or swerve and strike the entrance of the tunnel, killing the car's occupant. What should it do?

Media attention surrounding ethical decision making in automated vehicles often begins with a similar example as the one above, where a vehicle must choose between two crash alternatives with almost certain fatal outcomes (Achenbach 2015; McFarland 2015). Scientific literature on vehicle automation uses similar examples (Lin 2015; Goodall 2014; Terken 2015; Millar 2014; Sandberg and Bradshaw-Martin 2013). These scenarios are based on the well-known philosophical thought experiment known as the trolley problem. Introduced by Philippa Foot (1967), and later elaborated by Judith Jarvis Thomson (1985; 2008), the trolley problem originally described a tram traveling toward five people on a track. Those five face certain death unless a bystander pulls a switch, diverting the trolley onto a different track where it will kill one person. The thought experiment has been altered to describe vehicle automation by replacing the trolley with an automated vehicle, the bystander with the automated vehicle's pre-programmed software, and the track with a roadway that has some type of barrier to prevent simply driving onto the shoulder (e.g. a tall bridge, a tunnel, large trees on the shoulder, etc.).  Just as the original trolley problem inspired several variations, there are different versions using automated vehicles, some of which involve at least one choice that kills or injures the automated vehicle's passenger  (Bonnefon, Shariff, and Rahwan 2016; Moon et al. 2014).

In this paper, the term *trolley problem* is used to refer to both trolley and automated vehicle thought experiments.

*The Trolley Problem's Utility*

The trolley problem, including its variants, is useful for at least two reasons. First, it helps identify common responses or intuitions about the correct course of action, as well as areas of strong agreement or disagreement. By altering the trolley problem's rules while keeping the end result the same, philosophers can begin to explore the reasoning behind people's responses (and the philosophers' own intuitions), even if the respondents are unable to articulate their own reasoning.

While many find it permissible to pull the switch to kill one and save the five, those same people find it wrong to push a fat man onto the track to cause a derailment and meet the same objective. (The bystander cannot throw himself in front of the train, because he is not large enough to cause a derailment.) There are a few theories to explain this discrepancy, and by adjusting the scenario, one can begin to isolate the root cause. The scenario could instead have the bystander pull a switch which triggers a trap door that drops the fat man onto the tracks, so that the bystander does not have direct physical contact with the fat man. If the respondents generally find this action acceptable, then it is possible that the physical contact or proximity to the victim is a factor in the decision. (In surveys, respondents indicated that use of force rather than mere physical contact seems to make the action slightly less moral (Greene et al. 2009).) Studies using vehicle automation versions of the trolley problem have shown that survey respondents strongly disagree over how much risk a car's passenger should have to take on in order to protect pedestrians (Bonnefon, Shariff, and Rahwan 2016; Moon et al. 2014).

The second reason that the trolley problem and other thought experiments are helpful is that they serve as edge cases. Just as any good ethical theory should be able to respond reasonably in a thought experiment, so too should a sophisticated crash avoidance strategy have a defensible solution for these admittedly rare situations. Decisions involving life and death are among the most challenging a vehicle will face. A proposed ethical system for vehicles can be quickly checked against a range of hypothetical scenarios for unexpected or undesirable behavior. Suppose an ethical strategy calls for an automated vehicle to maximize global safety at all times. This strategy falls short in a scenario where an automated vehicle must choose between colliding with one of two motorcyclists, only one of which is wearing a helmet. An automated vehicle programmed to maximize safety would always select the helmeted motorcyclist, although many would find that unfair. This suggests that some information (weighing helmet vs. no helmet) should be excluded from an automated vehicle's decision making, or that there are shortcomings to a simple strategy of maximizing safety.

*The Trolley Problem's Shortcomings*

While research into automated vehicle ethics can start with the trolley problem, it should not end there. I suspect the trolley problem is so often employed by the media for four reasons: it is fairly well-known, it is a stark choice between a manageable number of alternatives, it has completely certain outcomes, and these outcomes have obvious moral implications. When the media refers to the trolley problem in the context of vehicle automation, they seem to use it as a stand-in for a range of more subtle ethical decisions an automated vehicle may face, many of which will have less obvious moral undertones, uncertain outcomes, and consequences that are not life-threatening.

This is a problem because critics of automated vehicle ethics can argue that any research into ethical decision making for automated vehicles is unnecessary or wasteful simply by attacking the trolley problem. The trolley problem is easy to attack. These are the most common criticisms of the trolley problem when applied to automated vehicles: it represents a false dilemma, it uses predetermined outcomes, it assumes perfect knowledge of its environment, and it has an unrealistic premise.

The trolley problem almost always involves two choices. This strikes many as unrealistic, and that any real pre-crash situation would have a range of alternatives. While the trolley problem is limited to two choices, this is primarily to facilitate discussion. In most driving, there may be an obviously safer third alternative, or there may be a third alternative that balances the danger among the affected parties. An automated vehicle in a tunnel trying to dodge a child may opt to swerve as much as possible without striking the wall, with the hope that the child moves out of the way just enough to survive, if not entirely avoid, the crash. This kind of probabilistic, risk balancing action is impossible in the trolley problem, where one must choose between two extreme actions, both of which always involve certain death.

Trolley problems also assume known outcomes, which critics are quick to dismiss as unrealistic—experts can confidently predict the outcomes of only the most catastrophic crashes (O'Donnell and Connor 1996; Kockelman and Kweon 2002), yet almost everything else in road safety is probabilistic. Fatality rates depend on such arbitrary inputs as whether a passenger is sober or drunk (twice as likely to die if drunk) , male or female (28% more likely to die if female), young or old (a 70-year-old is three times more likely to die than a 20-year-old) (Evans 2008). This is in sharp contrast with the trolley problem, where the result of pulling or not pulling the switch is always completely certain.

Sensor readings are also probabilistic. An automated vehicle operates using uncertain inputs. It may determine that a pedestrian is ahead on the sidewalk with 90% confidence, that there is a 70% chance she will run into the road based on her recent mannerisms and behavior, and that the automated vehicle is 98% certain that its blind spot is clear of other vehicles. Should it switch lanes? This requires straightforward risk calculations based on highly uncertain information. This is not easy. It requires defining magnitudes or "costs" of bad events. This is difficult to do in conversation or in a short article, and hence the need to simplify a discussion of automated vehicle ethics with an example like the trolley problem and its defined premise and deterministic outcomes.

Finally and most problematically, critics dismiss ethics generally because the trolley problem seems unrealistic, or at least highly improbable (Rose 2016). Most people have trouble recalling a driving incident where they had to make a decisions with what felt like genuine ethical considerations. Even if they can recall a decision requiring ethics, it is rarely something with fatal consequences like in the trolley problem. The vast majority of drivers will never be involved in a fatal collision, let alone one where they have time to consider and reflect on the correct course of action.

This is one area where it is important to distinguish between the trolley problem specifically and morally ambiguous scenarios generally. Drivers may feel like they have never made an ethical decision because they did not consider the ethics due to time constraints, or because they did not crash, yet any maneuver to avoid a jay-walking pedestrian in the roadway that puts the driver in slight danger is an ethical decision—the driver put herself at some risk, even a small amount, to protect a person to whom she owed no legal duty. This may have been an instinctual response from the driver, but in the days of vehicle automation, instinct will be replaced by decisions and logic encoded in software, sometimes programmed years before the crash.

The behavior of the vehicle can have an ethical component, even if the vehicle is not in immediate danger. The decisions of  how to position itself within a lane, how much buffer to provide a pedestrian, whether the buffer size should change based on a pedestrian's behavior or physical attributes, what type of headway to allow—these all carry some risk of crashing. When the answers to these questions are encoded in software, any imbalance in crash risk will be translated into an corresponding increase in actual crashes. While ethical decisions seem rare in daily driving, this is only true if one sees ethics in only life-or-death decisions as in the trolley problem. More subtle but just as difficult choices exist for almost all driving.

While the trolley problem is valuable in isolating people's intuitions about morally ambiguous crash decisions and stress testing ethical strategies, it represents a fairly narrow area of automated vehicle ethics and suffers from a perceived lack of realism. A better way of framing and studying automated vehicle ethics may be through the more applied field of risk management, discussed in the following section.

**Risk Management**

In common usage, the word "risk" refers to an unwanted event (which may or may not occur), the cause of that event, or its probability. For example, one might talk about the risk of lung cancer, the risk of smoking as it contributes to lung cancer, or the risk of shortening one's life by smoking of 50% (Hansson 2014). In formal risk analysis, the definition changes to the *expectation value* of an unwanted event which may or may not occur. With this definition, both the severity and the probability of the unwanted event are included in the definition.

The expectation value represents the product of two other values specific to that risk: its magnitude and its probability of occurrence. In most fields, the expectation value is expressed in terms of human life. A nuclear reactor that has a 0.01% chance of a meltdown occurring and which would cause 10,000 deaths is said to have an expectation value of 0.01% of 10,000 lives, or 1 human life. By calculating the expectation value of a nuclear reactor meltdown, and weighing it against the costs of various safety improvements, management can justify implementing (or not implementing) those improvements, provided that different metrics such as money and human life can be translated into directly comparable values.

The practice of calculating risks is generally referred to as risk analysis, while the decisions of how to allocate and balance these risks is referred to as risk management. While risk management has generally been applied to large scale technological, societal, or natural risks such as radiation exposure, terrorism, and flooding, the same strategies can be applied to smaller-scale, near-term problems such as automated vehicle decision making.

An example of risk management applied to vehicle automation was demonstrated in a patent awarded to Google entitled *Consideration of Risks in Active Sensing for an Autonomous Vehicle* (Teller and Lombrozo 2015). In the patent, the authors describe an automated vehicle stopped behind a large truck at an intersection. The truck is blocking the vehicle's view of the traffic signal. The vehicle has the option of moving into the left lane, which would give it a better view of the traffic signal and help determine whether the truck was simply waiting at a red light or was stopped for some other reason (e.g. broken down or making a delivery) at a green light. Moving into the left lane, however, carries some small risks such as a sensor failing to detect a car in the automated vehicle's blind spot, leading to a collision. An automated vehicle needs a way to determine if the benefits of moving into the left lane (better information about the traffic signal status and therefore the truck's intentions) outweigh the costs (a small risk of a crash).

The basic form of risk management consists of considering each potential outcome's magnitude and likelihood. Its magnitude is a measure of how severe an event would be should it occur, and can be expressed as either a positive or negative value. Although magnitude is typically expressed as the number of expected fatalities, it can also be converted into other values for comparing across categories, such as the value of $9.4 million to prevent a fatality recommended by the United States Department of Transportation (USDOT 2015b). Events that do not impact safety can also be assigned magnitudes—the USDOT recommends valuing the amount of travel time saved as $13.30 per person per hour (USDOT 2015a).

The second component of risk analysis is an event's likelihood of occurring, expressed as a probability. It is often derived from historical experiences, empirical data, models, or in some cases, educated guesses. Some analyses employ a third component: uncertainty of the likelihood value.

To calculate risk, an outcome's magnitude is multiplied by its likelihood. Suppose a safety improvement has a 1% chance of saving 200 lives. This improvement would have a value of preventing a fatality of 200 x 0.01 = 2. The cumulative risk for all outcomes for a single alternative can then be compared to other alternatives. For an automated vehicle deciding whether to go around the stopped truck, the risks may look like those shown in Figure 1, taken from the previously discussed patent (Teller and Lombrozo 2015) . The magnitudes in Figure 1 are merely examples, and may not represent those used in practice, if such a system is or were to be deployed at all. Example costs for crash injuries can instead be found in other sources (Blincoe et al. 2015).

Figure 1. Example of risk analysis for an automated vehicle, from a recent patent (Teller and Lombrozo 2015).

| Bad Event | Risk Magnitude | Probability (%) | Risk Penalty |
|---|---|---|---|
| getting hit by large truck | 5,000 | 0.01% | 0.5 |
| getting hit by an oncoming vehicle | 20,000 | 0.01 | 2 |
| getting hit from behind by vehicle (not shown) approaching in the left-hand lane 408 | 10,000 | 0.03% | 3 |
| hitting pedestrian who runs into the middle of the road | 100,000 | 0.001% | 1 |
| losing information that is provided by camera in current position | 10 | 10% | 1 |
| losing information that is provided by other sensor in current position | 2 | 25% | 0.5 |
| Interference with path planning involving right turn at traffic light 412 | 50 | 100% (if turn is planned)/0% (if no turn is planned) | 50/0 |

*Advantages of Risk Management*

There are several advantages to using risk management techniques to determine the best course of action. Unlike approaches that use machine learning and other black box artificial intelligence approaches, risk management is transparent and easily adjustable. In the event of a crash or near miss, investigators could request to see the vehicle's calculations, and trace back its logic to determine why it behaved in a certain way. For vehicles behaving unsafely or unexpectedly, their risk management settings can be adjusted. Probabilities of bad events can be updated either automatically or manually to reflect recent experiences, while magnitudes can also be adjusted to reflect society's preferences. These preferences may be explicit, for example, assigning equal values to individual human lives. These preferences could also be implicit, whereby magnitudes are calibrated to encourage the car to drive in a way that seems reasonable and predictable.

Another advantage of risk management is that, like the utilitarianism it is based on, it always recommends an action. Vehicle control systems that are based on deontological ethics must obey certain rules, and when these rules conflict, the vehicle may be forced to fall back on a less than ideal secondary approach. This problem was demonstrated in Isaac Asimov's short story *Runaround* (1942)*,* in which a robot programmed is programmed to follow two laws: obey orders and protect itself. The robot follows an order which simultaneously carries some unforeseen risk to its systems, and becomes stuck in a feedback loop. Without a fall back plan, an automated vehicle finding itself in a situation in which its programming does not recommend any action may come to a complete stop, which is often unsafe on high-speed roads. Risk management avoids this problem by always recommending some action, even if the recommendation is to hold course.

Similarly, in a deontological system, a vehicle may attempt a dangerous behavior in an attempt to avoid violating one of its programmed constraints. In the vehicle ethics system proposed by Gerdes and Thornton (2015), an automated vehicle is programmed to avoid pedestrians and cyclists first, other vehicles second, and objects third. An automated vehicle with this programming facing a possible crash may accelerate to avoid a crash with another vehicle, even if there is only a 1% chance of avoiding the collision, rather than decelerate and accept a low-speed and much less severe crash. Without some consideration of the probabilistic and uncertain nature of current and future conditions, a deontological car may undertake dangerous behaviors in order to maintain even a small chance of adhering to its rules.

*Shortcomings of Risk Management*

Risk management, especially classical risk management, has several disadvantages. Most risk management approaches consider all human life to be equivalent, with the value-of-a-statistical life disregarding precisely who is at risk and how. This seems fair and equitable at first, but fails when considering that not all participants benefit equally from the risk. An example of unequal benefit can be seen in a recent patent which describes an automated vehicle choosing to position itself laterally in a lane (Dolgov and Urmson 2014). The patents describes a situation where a vehicle in the middle lane of a freeway finds itself positioned between a large truck on its left and a small car on its right. In order to maximize net safety, it would position itself away from the large truck and closer to the small car, presumably because a crash with the small car would be less severe and safer overall. Over millions of miles, truck drivers would be expected to experience slightly fewer crashes than drivers of small cars, due to the lane positioning of the automated vehicle. This transfers risk from one party to another, without anyone's consent. While this may prove to be legal behavior, it seems unfair. The public may find the vehicle's behavior ultimately acceptable if it is accompanied by transparency in how the software determines and apportions risk, as well as a coherent justification for why the vehicle positions itself as it does.

One must also be cautious in applying the value of a statistical life to situations where death is fairly certain. Generally, the value of a statistical life is the amount of money one will spend to save a statistical life, not an actual one. The USDOT value of a statistical life does not imply that one can pay $9.4 million and then legally kill another. Likewise, most people do not use this value when deciding whether to rescue miners trapped underground. The value of a statistical life is used to prioritize safety improvements to minimize *anticipated* deaths, not actual ones. Risk management schemes should consider an exponential rather than linear relationship with probability, so that the value of a life increases dramatically the more likely it is that death would occur. Similar strategies that better reflect society's preferences would also be acceptable.

Finally, one must be careful to avoid unintentional discrimination. Some vehicles in collisions are known to inflict more damages than others. Two vehicles that inflict an unusually high amount of damage upon another are said to have poor crash compatibility (Gabler and Hollowell 2000). An automated vehicle may be programmed to keep a larger headway when following a vehicle with poor crash compatibility. Because vehicles vehicle mass is high correlated with crash severity, this behavior could eventually lead to disproportionately more crashes with smaller, cheaper cars. If these cheaper cars are predominately owned by low-income individuals, the automated vehicle's following behavior could be seen many to be discriminatory, even though it was intended only to improve safety. This is not to imply that using different following distances based on crash compatibility is itself unethical, but rather that these decisions should be considered from a moral perspective, and should use the language of ethics to justify these decisions.

Hansson describes three parties involved in any risk decision: the risk-exposed, the beneficiary, and the decision-maker. These roles often overlap, as the automated vehicle passenger who exposes herself to some risk while traveling is also benefitting by using the vehicle. Hansson proposes seven questions that address many of the ethical issues in risk management. Quoting directly from (Hansson 2007a):

1. To what extent do the risk-exposed benefit from the risk exposure?
2. Is the distribution of risks and benefits fair?
3. Can the distribution of risks and benefits be made less fair by redistribution or by compensation?
4. To what extent is the risk exposure decided by those who run the risk?
5. Do the risk-exposed have access to all relevant information about the risk?
6. Are there risk-exposed persons who cannot be informed or included in the decision process?
7. Does the decision-maker benefit from other people's risk exposure?

Translating these questions into vehicle automation can be challenging with shifting and overlapping roles of beneficiary, risk-exposed, and decision-maker. Still, these are important questions to answer when designing even seemingly straightforward automated driving functions, and the process of answering these questions can help in justifying the vehicle's actions after the fact.

**Conclusion**

Much of the discussion of automated vehicle ethics has focused on hypothetical dilemmas such as the trolley problem and its variants. These are useful for framing a discussion of ethics, identifying and classifying public response to these situation, and stress testing different ethical theories for automated vehicles. They are also easy to for critics to dismiss as unrealistic, and therefore unimportant and distracting. While trolley problems represent interesting and morally ambiguous situations automated vehicles may face, they are not the only problems with ethical implications. Vehicles must constantly assess the value of dangerous actions, and especially in crashes, must compare the values of different objects on the road. Decisions about how these values are formulated, and how the risks of driving are distributed to other road users, have ethical components. Risk management techniques can be used to quantify these probabilistic risks in a way that is transparent and adjustable. Care must be taken to apply ethical principles to risk management. A road user's exposure to risk should consider his vulnerability, voluntariness, and consent.

While the problem of integrating ethics into a vehicle's decision-making processes appears overwhelming, designing moral vehicles is a manageable problem. Strategies for how to distribute risk or ration benefits have been developed in many other areas including organ donation (Persad, Wertheimer, and Emanuel 2009), radiation exposure (Hansson 2007b), and industrial safety standards (Aven 2007). Ethicists can and should be consulted to ensure that the solutions, while imperfect and lacking consensus support, are justifiable using widely accepted ethical principles. Developers of automated vehicles should seek to learn from the experiences of experts in similar fields when developing ethical vehicles, and continue to study the ethical implications a vehicle's behavior.